\documentclass[aps,nofootinbib,showpacs,preprintnumbers,twocolumn,superscriptaddress]{revtex4}
\usepackage{graphicx}
\usepackage{dcolumn}
\usepackage{amsmath}

\newcommand{\MeV}{\mbox{\rm MeV}}

\newcommand{\Lqcd}{\Lambda_{\mbox{\rm\scriptsize QCD}}}
\newcommand{\lsim}[1]{
\setlength{\unitlength}{10pt}
\begin{picture}(1.4,1.)
\put(.7,-0.3){\makebox(0.0,1.)[t]{$<$}}
\put(.7,-0.3){\makebox(0.0,1.)[b]{$\sim$}}
\end{picture}#1}

\newcommand{\minimal}{\mbox{\rm\scriptsize min}}
\newcommand{\maximal}{\mbox{\rm\scriptsize max}}
\newcommand{\eff}{\mbox{\rm\scriptsize eff}}
\newcommand{\pert}{\mbox{\rm\scriptsize pe}}
\newcommand{\Npert}{\mbox{\rm\scriptsize Npe}}

\newcommand{\Mgluon}{m_A}
\newcommand{\Cgluon}{C_A}

\newcommand{\Cghost}{C_G}

\begin{document}

%\preprint{ZTF-03/05}

\title{Pseudoscalar $q\bar q$ mesons and
effective QCD coupling enhanced by 
$\langle A^2 \rangle$ condensate}

\author{Dalibor Kekez}
\affiliation{\footnotesize Rudjer Bo\v{s}kovi\'{c} Institute,
         P.O.B. 180, 10002 Zagreb, Croatia}

\author{Dubravko Klabu\v{c}ar}
\affiliation{\footnotesize Department of Physics, Faculty of Science, 
        Zagreb University, P.O.B. 331, 10002 Zagreb, Croatia}

\begin{abstract}
\rule{0ex}{3ex}
\noindent Recent developments provided evidence that 
the dimension 2 gluon condensate $\langle A^2 \rangle$
is important for the nonperturbative regime of Yang-Mills 
theories (quantized in the Landau gauge).
We show that it may be relevant for the Dyson-Schwinger
approach to QCD. In order that this approach leads to
a successful hadronic phenomenology, an enhancement of 
the effective quark-gluon interaction seems to be needed
at intermediate ($p^2 \sim 0.5$ GeV$^2$) momenta. 
It is shown that the gluon condensate $\langle A^2 \rangle$
provides such an enhancement. It is also shown that 
the resulting effective strong running coupling leads 
to the sufficiently strong dynamical chiral symmetry 
breaking and successful phenomenology at least in the 
light sector of pseudoscalar mesons.
\end{abstract}

\maketitle

\section{Introduction and survey}

Dyson-Schwinger (DS) equations provide a prominent 
approach to physics of strong interactions.  
One of its aspects, reviewed \cite{Alkofer:2000wg}
and exemplified 
recently by Refs. \cite{Alkofer:2000wg,Fischer:2003rp}, 
consists of {\it ab initio} studies of 
DS equations for Green's functions of QCD, typically 
in the Landau gauge (LG).
The other aspect consists of phenomenological 
DS studies (also typically in LG) of hadrons as quark bound 
states, but relying more on modeling; e.g., see a recent review 
\cite{Maris:2003vk}.
Such phenomenological studies have mostly 
been relying on the 
%{\it (renormalization-group-improved)}
rainbow-ladder approximation (RLA), 
where generation of dynamical chiral symmetry breaking (D$\chi$SB)
is well-understood 
\cite{Alkofer:2000wg,Maris:2003vk,Jain:qh,Maris:1997tm,Maris:1999nt}.
As it has been stressed in, e.g., 
Refs. \cite{Maris:2003vk,Holl:2003dq}, 
RLA is the leading-order term of a procedure 
\cite{Bender:1996bb,Bender:2002as} that can be systematically  
improved towards less severe truncations of DS equations. 
This general procedure 
provides a means to identify {\it a priori} the channels
in which RLA is likely to work well \cite{Bender:2002as}. 
Pseudoscalar mesons are the most notable among those channels
because of the correct chiral QCD behavior due to D$\chi$SB 
and Goldstone theorem, since (almost) massless pseudoscalar 
mesons are reproduced in the (vicinity of) chiral limit not 
only by the exact QCD treatment but also by all consistent 
truncations such as RLA \cite{Maris:1997hd}.

Consistent RLA implies 
\cite{Alkofer:2000wg,Maris:2003vk,Jain:qh,Maris:1997tm,Maris:1999nt} 
that {\it Ans\" atze} of the form 
\begin{equation}
[K(p)]_{ef}^{hg} = - 4\pi\alpha_{\mbox{\rm\scriptsize eff}}(p^2) \,
       D_{\mu\nu}^{ab}(p)_{0} \,
[\frac{\lambda^a}{2}\,\gamma_{\mu}]_{eg} 
[\frac{\lambda^b}{2}\,\gamma_{\nu}]_{hf}
\label{RLAkernel}
\end{equation}
must be used for the interactions between quarks in both the gap equation 
$S^{-1} = S^{-1}_{0} - \Sigma$ for the full quark propagator $S$ 
($S_{0}$ is the {\it free} one) and the Bethe-Salpeter (BS) equation 
for the meson ($\tt M$) bound-state vertex $\Gamma_{\tt M}$; i.e., 
\begin{equation}
[\Sigma]_{ef} = \int S_{gh} [K]_{ef}^{hg}  \, \, , \,
[\Gamma_{\tt M}]_{ef} = \int [S \Gamma_{\tt M} S ]_{gh} [K]_{ef}^{hg} \, \, ,
\label{schematically}
\end{equation}
where, writing {\it schematically}, integrations are meant over loop 
momenta while $e,f,g,h$ in Eqs. (\ref{RLAkernel})-(\ref{schematically})
represent spinor, color and flavor indices \cite{Maris:2003vk}. 
In LG, the {\it free} gluon propagator $D_{\mu\nu}^{ab}(p)_{0} \equiv 
\delta^{ab}(\delta_{\mu\nu}-p_\mu p_\nu/p^2)/p^2$.

\begin{figure}[t]
\centerline{\includegraphics[height=66mm,angle=0]{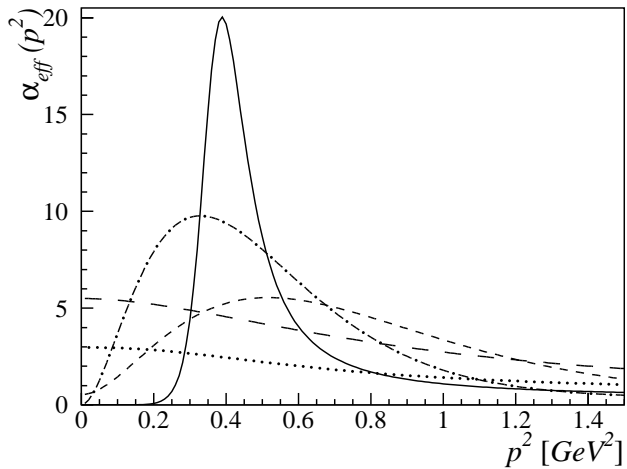}}
\caption{The momentum dependence of various strong running couplings
mentioned in the text. JM \cite{Jain:qh} and MRT
\cite{Maris:1999nt,Maris:2003vk} $\alpha_{\eff}(p^2)$
are depicted by, respectively, dashed and dash-dotted curves.
The effective coupling (\ref{ourAlpha_eff}) proposed and analyzed
in the present paper is depicted by the solid curve, and
$\alpha_{\mbox{\rm\scriptsize s}}(p^2)$~(\ref{Alkofalpha})
of Fischer and Alkofer \cite{Fischer:2003rp} (fit A)
by the dotted curve.  The long-dashed curve is 
the fit (extrapolated all the way to $p^2=0$) of the 
lattice results of Ref. \cite{Bloch:2002we}. }
\label{Fig1}
\end{figure}

In Eq. (\ref{RLAkernel}), $\alpha_{\mbox{\rm\scriptsize eff}}(p^2)$ 
is an effective running coupling. It is only partially known 
from the fact that at large 
spacelike momenta (our convention is $p^2>0$ for spacelike $p$), 
$\alpha_{\mbox{\rm\scriptsize eff}}(p^2)$ must reduce to 
$\alpha_{\pert}(p^2)$, the well-known running coupling of 
perturbative QCD.
However, for $p^2 \lsim 1$ GeV$^2$, where nonperturbative QCD 
applies, the interaction is still not known. Thus, 
in phenomenological DS studies, $\alpha_{\eff}(p^2)$ must 
be modeled for $p^2 \lsim 1$ GeV$^2$ - e.g., see Refs.
\cite{Jain:qh,Maris:1997tm,Maris:1999nt,Alkofer:2000wg,Maris:2003vk}.
There one can see that phenomenologically most successful of those 
modeled interactions have a rather large bump at intermediate 
momenta, around $p^2 \sim 0.5$ GeV$^2$ 
(e.g., in Fig. \ref{Fig1} see $\alpha_{\eff}(p^2)$ used by Jain and 
Munczek (JM) \cite{Jain:qh} and by Maris, Roberts and Tandy (MRT)
\cite{Maris:1997tm,Maris:1999nt,Maris:2003vk}).
In any case, 
successful DS phenomenology demands that the modeled part of 
the interaction (\ref{RLAkernel}) be fairly strong; 
regardless of details of the form of the interaction, its 
{\it integrated strength} (for $p^2 \lsim 1$ GeV$^2$) must be fairly high 
to achieve acceptable description of hadrons, notably mass spectra 
and D$\chi$SB \cite{Maris:2003vk,Maris:2002mt}. 
Theoretical explanations of the origin of so strong 
nonperturbative part of the phenomenologically required interaction 
are obviously much needed, either from the {\it ab initio} DS 
studies or from somewhere outside the DS approach. 
This is the main motivation for the present paper.

The {\it ab initio} DS studies 
showed that, in LG, the effects of ghosts are 
crucial for the intermediate-momenta enhancement of the effective 
quark-gluon interaction. 
This is obvious in the expression for the strong running coupling 
$\alpha_{\mbox{\rm\scriptsize s}}(p^2)$ in these LG studies
\cite{Alkofer:2000wg,Fischer:2003rp},
\begin{equation}
\alpha_{\mbox{\rm\scriptsize s}}(p^2)
=
\alpha_{\mbox{\rm\scriptsize s}}(\mu^2) \, Z(p^2) \, G(p^2)^2 
\, ,
\label{Alkofalpha}
\end{equation}
where $\alpha_{\mbox{\rm\scriptsize s}}(\mu^2) = g^2/4\pi$ and 
$Z(\mu^2) G(\mu^2)^2 = 1$ at the renormalization point $p^2 = \mu^2$.
The ghost and gluon renormalization functions $G(p^2)$ and $Z(p^2)$
define the full ghost propagator $D_G^{ab}(p) = - \delta^{ab} G(p^2)/p^2$
and the full LG gluon propagator 
$D_{\mu\nu}^{ab}(p) = Z(p^2)\, D_{\mu\nu}^{ab}(p)_{0}$.

While the {\it ab initio} DS studies 
\cite{Alkofer:2000wg,Fischer:2003rp} 
do find significant enhancement of $\alpha_{\mbox{\rm\scriptsize s}}(p^2)$, 
Eq. (\ref{Alkofalpha}), until recently this seemed still not enough 
to yield a sufficiently strong D$\chi$SB 
(e.g., see Sec.~5.3 in Ref.~\cite{Alkofer:2000wg})
and a successful phenomenology. Nevertheless, 
going beyond the ladder truncation and so getting
additional interaction strength from dressed vertices,
for carefully 
constructed dressed quark-gluon vertex {\it Ans\" atze},
Fischer and Alkofer \cite{Fischer:2003rp} have recently 
obtained good results for 
%dynamically generated 
constituent quark masses and pion decay constant $f_\pi$,
although not simultaneously also for the chiral quark-antiquark
$\langle \bar{q}q \rangle$ condensate, which then came out 
somewhat too large. Thus, 
the overall situation is that there is progress 
\cite{Fischer:2003rp,Maris:2002mt}, 
but that further investigation and elucidation of 
the origin of phenomenologically successful effective interaction 
kernels remains one of primary challenges in 
%contemporary 
DS studies \cite{Maris:2003vk}. 
Here, we will point out that such an interaction kernel for DS studies 
results from 
combining the DS ideas on $\alpha_{\mbox{\rm\scriptsize s}}(p^2)$
of the form (\ref{Alkofalpha}) \cite{Alkofer:2000wg,Fischer:2003rp} 
and the ideas 
\cite{Boucaud:2000nd,Gubarev:2000eu,Gubarev:2000nz,Kondo:2001nq,Celenza:1986th} 
on the dimension 2 gluon condensate 
$\langle A_\mu^a A^{a\mu} \rangle \equiv \langle A^2 \rangle$
{\it in LG}.

\section{Condensates in gluon and ghost propagators}

Already a long time ago 
Refs. \cite{Lavelle:eg,Lavelle:xg,Ahlbach:ws,Lavelle:yh}
found in the operator product expansion (OPE) 
the $\langle A^2 \rangle$-contributions to QCD propagators,
recently confirmed by Kondo \cite{Kondo:2001nq}.
For LG (adopted throughout this paper), 
number of colors $N_c = 3$ and space-time dimensions $D=4$, 
their results for gluon and ghost propagators amount to
\begin{equation}
Z(p^2) = \frac{1}{1 + \frac{m_A^2}{p^2} + 
\frac{{\cal O}_A(1/p^2)}{p^2} }
\, ,
\label{KondoZ}
\end{equation}
\begin{equation}
G(p^2) = \frac{1}{1 + \frac{m_G^2}{p^2} + 
\frac{{\cal O}_G(1/p^2)}{p^2} }
\, ,
\label{KondoG}
\end{equation}
\begin{equation}
m_A^2 = \frac{3}{32} \,\,  g^2 \langle A^2 \rangle = - m_G^2~,
\label{gluonMass}
\end{equation}
where $m_A$ and $m_G$ are, respectively, dynamically generated
effective gluon ($A$) and ghost ($G$) mass.
The later references \cite{Kondo:2001tm,Kondo:2002xn}
also worked out the perturbative QCD corrections inducing
the logarithmic $p^2$-dependence of these dynamically generated
masses, i.e. $m_A^2(p^2)$, to which we will return and comment on 
in the next section.

These now well-established propagator contributions 
(\ref{KondoZ})-(\ref{gluonMass})
then suggest the importance of $\langle A^2 \rangle$
for the DS approach to hadrons, where propagators, usually in LG,
are used to get solutions for quark bound states
and calculate observable quantities.
Notably, see Ref. \cite{Maris:1997hd} for gauge-parameter independent
expressions for $f_\pi$ and 
a generalization of the Gell-Mann-Oakes-Renner relation (GMOR)
that demonstrates gauge-parameter independence of the meson mass.
Still, how can $\langle A^2 \rangle$
influence these observable quantities, since 
this condensate is {\it not} gauge-invariant?
It turns out \cite{Gubarev:2000eu,Gubarev:2000nz,Dudal:2002xe}
that in LG, $\langle A^2 \rangle$
equals a non-local, but gauge-invariant quantity: the
minimal (with respect to the choice of gauge)
value of $A_{\mu}^{a}A^{\mu a}$
integrated over the space-time, indicating that
$\langle A^2 \rangle$ in LG may have a physical meaning. 
Outside LG, besides $\langle A^2 \rangle$ other (ghost) 
condensates of dimension 2 appear \cite{Kondo:2001nq}.
They very likely cancel the variation which $\langle A^2 \rangle$
suffers in going to another gauge, since the
physics behind all these different dimension 2 
condensates in different gauges must be the same: gluon-ghost
condensation lowers the QCD vacuum energy $E$, which is a
physical, gauge-invariant quantity, to a stable (``$E<0$'')
vacuum \cite{Dudal:2002xe}.

For $g^2 \langle A^2 \rangle$, LG lattice studies of 
Boucaud {\it et al.} \cite{Boucaud:2000nd} yield the value $2.76$ GeV$^2$,
compatible with the bound resulting from the discussions 
of Gubarev {\it et al.} \cite{Gubarev:2000eu,Gubarev:2000nz} 
on the physical meaning of $\langle A^2 \rangle$ and its 
importance for confinement. This value 
gives (\ref{gluonMass}) $m_A = 0.845$ GeV, which will 
turn out to be a very good initial estimate for $m_{A,G}$.

As for the contributions 
${\cal O}_A(1/p^2)$ and ${\cal O}_G(1/p^2)$
in Eqs. (\ref{KondoZ}) and (\ref{KondoG}), 
one expects a prominent role of the dimension 4 gluon condensate
$\langle F_{\mu\nu}^a F^{a\mu\nu} \rangle \equiv \langle F^2 \rangle$,
which, contrary to $\langle A^2 \rangle$, 
is gauge invariant \cite{Shifman:bx}.
Refs. \cite{Lavelle:xg,Ahlbach:ws} showed that 
the OPE contributions of dimension 4 condensates 
were far more complicated \cite{Lavelle:yh} than found 
previously \cite{Lavelle:eg}: not only many 
kinds of condensates contributed to terms $\propto 1/p^2$, but 
for many of them (gauge-dependent gluon, ghost and mixed ones) 
there has been no assignments of any kind of values yet.
Terms $\propto (1/p^2)^n$ $(n>1)$ were not considered at all. 
Thus, at this point, the only practical approach is that 
the contributions 
${\cal O}_A(1/p^2)$ and ${\cal O}_G(1/p^2)$
in Eqs. (\ref{KondoZ}) and (\ref{KondoG})
are approximated by the terms $\propto 1/p^2$ and parametrized, i.e.,
\begin{equation}
{\cal O}_A(1/p^2) \approx \frac{C_A}{p^2} \,\,\,  , \qquad
{\cal O}_G(1/p^2) \approx  \frac{C_G}{p^2} \,\,\,  ,
\label{calOs}
\end{equation}
where both $C_A$ and $C_G$ are in principle free parameters to be
fixed by phenomenology. Still, we should mention that the effective 
gluon propagator advocated by Lavelle \cite{Lavelle:ve} would indicate
$C_A \approx (0.640 \,\, {\rm GeV})^4$ for the following reason:
for 
LG and $D=4$, the contribution which this gluon 
propagator receives from the so-called ``pinch diagrams'' vanishes, 
and its \cite{Lavelle:ve} ${\cal O}_A(1/p^2)$ 
contribution
\begin{equation}
\Pi_A^{\langle F^2 \rangle}(p^2) 
      = 
 \frac{34 N_c \pi \alpha_s \langle F^2 \rangle}{9(N_c^2 - 1)p^2}
      =
 \frac{(0.640 \,\, {\rm GeV})^4}{p^2}
\label{PiLavelle91}
\end{equation}
stems entirely from the gluon polarization function in Ref.  
\cite{Lavelle:xg}, provided one invokes some fairly plausible
assumptions, like using equations of motion, to eliminate 
all condensates except $\langle F^2 \rangle$.
(The quark condensate $\langle {\bar q}q \rangle$ could
also be neglected \cite{Lavelle:ve}.)
Since Ref. \cite{Ioffe:2002be} indicates that the true value 
of $\alpha_s \langle F^2 \rangle$ is still rather uncertain, 
and since Refs. \cite{Ahlbach:ws,Lavelle:yh} make clear that 
Lavelle's \cite{Lavelle:ve} propagator misses some (unknown) 
three- and four-gluon contributions, we do not attach too much 
importance to the precise value $C_A = (0.640 \,\, {\rm GeV})^4$ 
\cite{Lavelle:ve,Shifman:bx} in Eq. (\ref{PiLavelle91}), 
but just use it as an inspired initial estimate. 
Fortunately, the corresponding variations of $C_A$ still
permit good phenomenological fits, since we will find below 
that our results are not very sensitive to $C_A$. 

There is no similar estimate for $C_G$,
but one may suppose that it would not differ from $C_A$ by orders of 
magnitude. We thus try $C_G = C_A = (0.640 \,\, {\rm GeV})^4$ as an
initial guess. It will turn out, {\it a posteriori}, that this value 
of $C_G$ leads to a remarkably good fit to phenomenology.

\section{Coupling enhanced by the gluon condensates}

Having set the stage, we are now ready to propose that 
$m_A^2 = - m_G^2 \propto \langle A^2 \rangle$ leads to 
the enhancement of $\alpha_{\eff}(p^2)$ at intermediate $p^2$.
To derive the running coupling exhibiting this property, let 
us first recall the aforementioned perturbative corrections to 
OPE results (\ref{KondoZ}), (\ref{KondoG})
and (\ref{gluonMass}). In Eqs.~(\ref{KondoZ}), (\ref{KondoG}) and
(\ref{calOs}), gluon and ghost renormalization functions are 
parametrized by the coefficients $m_A$, $C_A$, and $C_G$, 
which are constants at the tree level but develop momentum 
dependence through the perturbative corrections.
To see this, we note that the generic forms of the 
ghost and gluon renormalization functions including 
OPE contributions {\it and} perturbative QCD corrections
\cite{Boucaud:2001st,Kondo:2001tm,Kondo:2002xn}
can be written as 
\begin{eqnarray}
Z(p^2)
=
\frac{1}{r_0^A(p^2) +\frac{r_2^A(p^2)}{p^2} + 
\frac{r_4^A(p^2)}{p^4} + \dots},
\label{genericGluonRenormalizationFunction}
\\ 
G(p^2)
=
\frac{1}{r_0^G(p^2) +\frac{r_2^G(p^2)}{p^2} + 
\frac{r_4^G(p^2)}{p^4} + \dots},
\label{genericGhostRenormalizationFunction}
\\ \nonumber
\end{eqnarray}
where $r_0^A(p^2)$, $r_2^A(p^2)$, $r_4^A(p^2)$, etc., 
are the terms of mass-dimension 0, 2, 4, etc.,
for the gluon case, and 
$r_0^G(p^2)$, $r_2^G(p^2)$, $r_4^G(p^2)$, etc.,
are the terms of mass-dimension 0, 2, 4, etc.,
for the ghost case.
For example, the terms of the dimension zero, up to one loop, are
\begin{equation}
r_0^A(p^2) =
\left(\frac{\alpha_{\mbox{\rm\scriptsize pe}}(p^2)}
     {\alpha_{\mbox{\rm\scriptsize pe}}(\mu^2)}
\right)^{\gamma} \,\, , 
\label{dim0coeffA}
\end{equation}
\begin{equation}
r_0^G(p^2) =
\left(
\frac{\alpha_{\mbox{\rm\scriptsize pe}}(p^2)}
     {\alpha_{\mbox{\rm\scriptsize pe}}(\mu^2)}
\right)^{\delta} \,\, ,
\label{dim0coeffG}
\end{equation}
where $\gamma$ and $\delta$ are, respectively, gluon and ghost
anomalous dimensions. The perturbative corrections for the 
Wilson coefficients of dimension 2
% $\langle A^2\rangle$ condensate
have also been 
calculated for the pure Yang-Mills case 
by Boucaud {\em et al.} \cite{Boucaud:2001st} (for the gluon propagator) 
and Kondo and collaborators \cite{Kondo:2001tm,Kondo:2002xn} 
(for both the gluon and ghost propagators),
and since they imply that
\begin{equation}
\frac{r_2^A(p^2)}{r_0^A(p^2)} = - \frac{r_2^G(p^2)}{r_0^G(p^2)}
\equiv  m_A^2(p^2) = - m_G^2(p^2) \,\, ,
\label{p2depMassA}
\end{equation}
we write the renormalization functions as
\begin{eqnarray}
Z(p^2)
=
\left(
\frac{\alpha_{\mbox{\rm\scriptsize pe}}(p^2)}
     {\alpha_{\mbox{\rm\scriptsize pe}}(\mu^2)}
\right)^{-\gamma}
\frac{1}{1 +\frac{\Mgluon^2(p^2)}{p^2} + \frac{\Cgluon(p^2)}{p^4} + \dots},
\label{GluonRenormalizationFunction}
\\
G(p^2)
=
\left(
\frac{\alpha_{\mbox{\rm\scriptsize pe}}(p^2)}
     {\alpha_{\mbox{\rm\scriptsize pe}}(\mu^2)}
\right)^{-\delta}
\frac{1}{1 -\frac{\Mgluon^2(p^2)}{p^2} + \frac{\Cghost(p^2)}{p^4} + \dots}.
\label{GhostRenormalizationFunction}
\\ \nonumber
\end{eqnarray}
The perturbative corrections for the Wilson coefficients 
of dimension four and higher have not been calculated yet,
but we introduced also the notation
\begin{equation}
\frac{r_4^A(p^2)}{r_0^A(p^2)} = C_A(p^2) \,\, , \qquad 
\frac{r_4^G(p^2)}{r_0^G(p^2)} = C_G(p^2) \,\, ,
\end{equation}
to point out the correspondence of Eqs. 
(\ref{GluonRenormalizationFunction}) and 
(\ref{GhostRenormalizationFunction})
with the relations (\ref{KondoZ})-(\ref{gluonMass}) 
and (\ref{calOs}).
The latter differ from the former just by the absence of the 
slowly varying logarithmic $p^2$-dependence in $m_A(p^2)$, $C_A(p^2)$, 
$C_G(p^2)$ and $\alpha_{\mbox{\rm\scriptsize pe}}(p^2)$,
and of the dots denoting terms of dimension larger than four.
 
Regarding the perturbatively generated prefactors
($\alpha_{\mbox{\rm\scriptsize pe}}(p^2)/
\alpha_{\mbox{\rm\scriptsize pe}}(\mu^2)$ 
to the powers of $-\gamma$ and $-\delta$),
the forms (\ref{GluonRenormalizationFunction}) and 
(\ref{GhostRenormalizationFunction}) are consistent with the 
corresponding forms given by Eqs. (12) and (13) and also (14) in Ref. 
\cite{Fischer:2002hn} and by Eqs. (41) in Ref. \cite{Fischer:2003rp}. 
For $N_c$ colors and $N_f$ quark flavors, the anomalous dimensions 
of the gluon and ghost propagator are respectively given by $\gamma 
= (-13N_c+4N_f)/(22N_c-4N_f)$ and $ \delta = - 9N_c/(44N_c-8N_f)$ 
(see, e.g., Ref.~\cite{Fischer:2003rp}). This ensures 
$\gamma + 2 \delta = - 1$. The definition of the strong running 
coupling constant, Eq.~(\ref{Alkofalpha}), together with 
Eqs.~(\ref{GluonRenormalizationFunction}) and
(\ref{GhostRenormalizationFunction}), thus gives
\begin{eqnarray}
\alpha_s(p^2)
=
\alpha_{\mbox{\rm\scriptsize pe}}(p^2)
\frac{1}{1 +\frac{\Mgluon^2(p^2)}{p^2} + \frac{\Cgluon(p^2)}{p^4} + \dots  }
\nonumber \\ \times
\left(\frac{1}{1 -\frac{\Mgluon^2(p^2)}{p^2} + \frac{\Cghost(p^2)}{p^4} + \dots }\right)^2~.
\end{eqnarray}

\noindent Neglecting the $p^2$--dependence in the coefficients $\Mgluon$,
$\Cgluon$, and $\Cghost$, as well as the higher terms in the denominators,
we finally get
\begin{eqnarray}
\alpha_s(p^2) \approx 
\alpha_{\mbox{\rm\scriptsize pe}}(p^2)
\frac{1}{1 +\frac{\Mgluon^2}{p^2} + \frac{\Cgluon}{p^4}}
\left(\frac{1}{1 -\frac{\Mgluon^2}{p^2} + \frac{\Cghost}{p^4}}\right)^2
\nonumber   \\
    \label{ourAlpha_eff}      \\
\equiv 
\alpha_{\eff}(p^2)
= \alpha_{\mbox{\rm\scriptsize pe}}(p^2) \,
Z^{\Npert}(p^2) \, G^{\Npert}(p^2)^2~, \quad   \nonumber
\end{eqnarray}
depicted in Fig. \ref{Fig1} by the solid line [for the parameter
values (\ref{StandardParameterSet-new}), discussed below].
The suggestive abbreviations 
%\begin{eqnarray}
%Z^{\Npert}(p^2)&=&\frac{1}{1
%                   +\frac{m_A^2 }{p^2}
%                   + \frac{C_A}{p^4}}~
%\label{ZOPE}
%\\
%G^{\Npert}(p^2)&=&\frac{1}{1
%                   -\frac{m_A^2}{p^2}
%                   + \frac{C_G}{p^4}}~
%\label{ZGOPE}
%\end{eqnarray}
\begin{equation}
Z^{\Npert}(p^2) = \frac{1}{1
                   +\frac{m_A^2 }{p^2}
                   + \frac{C_A}{p^4}}~
\label{ZOPE}
\end{equation}
and
\begin{equation}
G^{\Npert}(p^2) = \frac{1}{1
                   -\frac{m_A^2}{p^2}
                   + \frac{C_G}{p^4}}~ 
\label{ZGOPE}
\end{equation}
for the factors giving the deviation of Eq. (\ref{ourAlpha_eff})
from the perturbative coupling 
$\alpha_{\mbox{\rm\scriptsize pe}}(p^2)$, stress that
our approximations amount to assuming that nonperturbative 
($\Npert$) effects are given by the OPE-based results of Refs. 
\cite{Lavelle:eg,Lavelle:xg,Ahlbach:ws,Lavelle:yh,Kondo:2001nq}
which in our present case boil down to 
Eqs.  (\ref{KondoZ})-(\ref{gluonMass}),
and by the parametrization (\ref{calOs}). 

\begin{figure}[b]
\centerline{\includegraphics[height=66mm,angle=0]{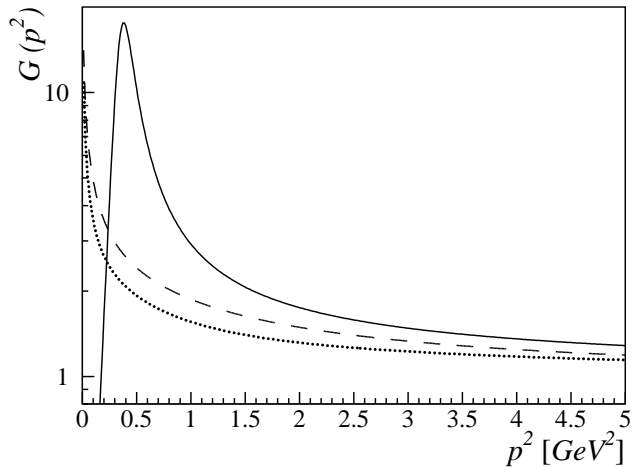}}
\caption{The momentum dependence of the ghost renormalization
function $G(p^2)$. The solid curve is our result 
(\ref{GhostRenormalizationFunction-approx}). 
The dotted curve is a result of the {\it ab initio} DS study 
\cite{Fischer:2003rp}. Concretely, it depicts the fit A in Eq. (41)
of Ref. \cite{Fischer:2003rp} evaluated at $\mu = 5$ GeV, for which 
value the dotted curve agrees rather well
(in the displayed momentum range) with the long-dashed curve,
representing Eq. (4) of the lattice Ref. \cite{Bloch:2002we},
renormalized at $\mu = 5$ GeV.
Our result (\ref{GhostRenormalizationFunction-approx}) is 
therefore plotted for the same value of $\mu$.  
 }
\label{FigG}
\end{figure}

\begin{figure}[t]
\centerline{\includegraphics[height=66mm,angle=0]{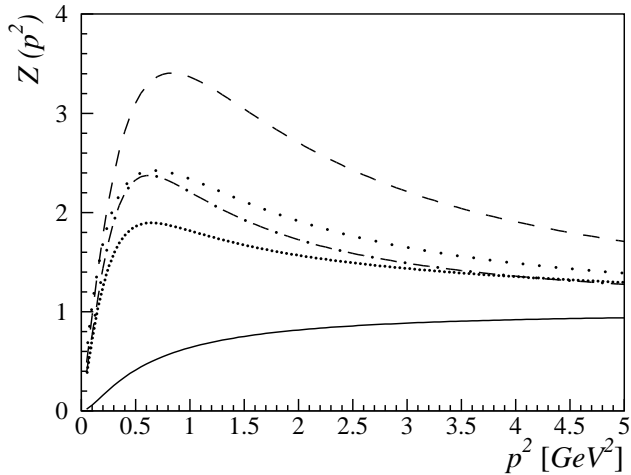}}
\caption{The momentum dependence of the gluon renormalization
function $Z(p^2)$.  The solid curve is our result 
(\ref{GluonRenormalizationFunction-approx}). 
The densely-dotted curve is the {\it ab initio} DS result 
(fit A) from Refs.  \cite{Fischer:2003rp,Alkofer:2002aa}. 
As in Fig. \ref{FigG}, both the {\it ab initio} DS result and 
our Eq.  (\ref{GluonRenormalizationFunction-approx}) are 
plotted for $\mu = 5$ GeV.  
The other curves pertain to some recent lattice results
\cite{Leinweber:1998uu,Iida:2003xe,BBLWZ00,Bloch:2002we}. 
The ones that agree with the {\it ab initio} DS result 
\cite{Fischer:2003rp,Alkofer:2002aa} very well, are those of 
Leinweber {\it et al.} \cite{Leinweber:1998uu} (their Eq. 
(5.14), depicted here by the sparsely-dotted curve), and 
Eq. (3) of Iida {\it et al.} \cite{Iida:2003xe} (dash-dotted curve) 
fitting the lattice data of Refs. \cite{BBLWZ00}. 
The third lattice result for $Z(p^2)$ is displayed by the long-dashed 
curve. It corresponds to Eq. (3) of Ref. \cite{Bloch:2002we}, which
is the fit (extrapolated all the way to $p^2=0$) to the lattice data
of Ref. \cite{Bloch:2002we}. For all lattice results,
the renormalization condition $Z(\mu^2)=1$ is imposed
at $\mu = 5$ GeV.  }
\label{FigZ}
\end{figure}

Our final expression (\ref{ourAlpha_eff}) for the running coupling, 
to be used in phenomenological calculations below, does not depend 
on the renormalization scale $\mu^2$ explicitly, although 
$Z(p^2)$ and $G(p^2)$, appearing in the intermediate steps, still do. 
Namely, the same approximation applied to the gluon and ghost 
renormalization functions (\ref{GluonRenormalizationFunction})
and (\ref{GhostRenormalizationFunction}), gives
\begin{eqnarray}
Z(p^2)
=
\left(
\frac{\alpha_{\mbox{\rm\scriptsize pe}}(p^2)}
     {\alpha_{\mbox{\rm\scriptsize pe}}(\mu^2)}
\right)^{-\gamma}
\frac{1}{1 +\frac{\Mgluon^2}{p^2} + \frac{\Cgluon}{p^4} }~,
\label{GluonRenormalizationFunction-approx}
\\
G(p^2)
=
\left(
\frac{\alpha_{\mbox{\rm\scriptsize pe}}(p^2)}
     {\alpha_{\mbox{\rm\scriptsize pe}}(\mu^2)}
\right)^{-\delta}
\frac{1}{1 -\frac{\Mgluon^2}{p^2} + \frac{\Cghost}{p^4} }~,
\label{GhostRenormalizationFunction-approx}
\end{eqnarray}
plotted in Figs. \ref{FigG} and \ref{FigZ} as our $G(p^2)$ and $Z(p^2)$ 
[again for the parameter values (\ref{StandardParameterSet-new})].

In our figures we also plot some results of lattice
and {\it ab initio} DS studies.
Namely, before we turn to our main goal, i.e., exploring whether 
our running coupling (\ref{ourAlpha_eff}) leads to successful 
phenomenology when used in quark gap DS and bound-state BS 
equations through Eq. (\ref{RLAkernel}), we will first comment
on the comparison of our gluon and ghost renormalization functions
with some other results.  In particular, recent T\"{u}bingen  
results of SU(2) lattice gauge simulations
\cite{Bloch:2002we,SaoCarlos,Bloch:2003sk} and of 
{\it ab initio} DS studies \cite{Fischer:2003rp,Alkofer:2002aa}
agree with each other well for $G(p^2)$ and quite reasonably
for $Z(p^2)$. 
This is seen in Figs. \ref{FigG} and \ref{FigZ}.
(Also, the results in most recent T\"{u}bingen lattice
reference \cite{Bloch:2003sk}, which however does not give the 
corresponding fitting formulas, are after proper renormalization
somewhat lower than the results of Ref. \cite{Bloch:2002we} plotted
in Figs. \ref{FigG} and \ref{FigZ}, and thus agree                        
somewhat better with the displayed {\it ab initio} DS results
\cite{Fischer:2003rp,Alkofer:2002aa}.)
As can be seen in Fig. \ref{FigZ}, there are some other
lattice results \cite{Leinweber:1998uu,Iida:2003xe,BBLWZ00}
for the gluon renormalization function which agree with $Z(p^2)$
from DS approach \cite{Fischer:2003rp,Alkofer:2002aa} even better,
but they do not give the ghost renormalization function.
Thus, T\"{u}bingen lattice results 
\cite{Bloch:2002we,SaoCarlos,Bloch:2003sk} are presently of 
particular interest, because they give both $Z(p^2)$ and $G(p^2)$.
Admittedly, there is a caveat: while the {\it ab initio} DS
studies \cite{Fischer:2003rp,Alkofer:2002aa} do not have problems
with reaching low momenta and in fact make strong statements
about the asymptotic behavior in the $p^2 \to 0$ limit, the
lattice data \cite{Bloch:2002we} do not reach very low momenta.
The lowest data point for $G(p)$ as well as $\alpha(p)$ in
Ref. \cite{Bloch:2002we} is at $p^2 \sim 0.36$ GeV$^2$, so that
one must keep in mind that for lower $p^2$, the corresponding
lattice-data-fitting curves in Figs. \ref{FigG} and \ref{FigZ},
and therefore also in Fig. \ref{Fig1}, are just extrapolations.
Nevertheless, presently most important is that comparing the 
long-dashed and dotted curves in Fig. \ref{Fig1} shows that
the respective running couplings (\ref{Alkofalpha}) following
from these ``lattice" \cite{Bloch:2002we,SaoCarlos,Bloch:2003sk}
and ``{\it ab initio} DS" \cite{Fischer:2003rp,Alkofer:2002aa}
renormalization functions typically do not differ by more than
the factor of two.
Thus, in spite of the mentioned caveat, these lattice results
\cite{Bloch:2002we,SaoCarlos,Bloch:2003sk} and the aforementioned
D$\chi$SB scenario of Fischer and Alkofer \cite{Fischer:2003rp}
support each other.
Still, the behavior of QCD propagators, especially the ghost ones,
and the resulting running coupling,
is not a closed issue yet, so that the presently proposed scenario
should also be considered although it is not supported by lattice
results.

The examples \cite{Boucaud:2003xi,Boucaud:2002fx,Suman:1995zg,Suman:be}
of lattice results differing from the T\"ubingen lattice
\cite{Bloch:2002we,Bloch:2003sk} and {\it ab initio} DS results 
\cite{Fischer:2003rp,Alkofer:2002aa} are not only the relatively old 
ones such as those of Suman and Schilling on the ghost renormalization 
function (which abruptly falls for the very smallest probed momenta, 
possibly indicating the infrared vanishing behavior) 
\cite{Suman:1995zg,Suman:be}, but also some of the most recent ones, 
such as Ref. \cite{Boucaud:2003xi}, where Landau gauge 
lattice calculations give the strong running coupling which, 
supposedly due to instanton effects, decreases{\footnote{Their 
strong running coupling becomes roughly 0.1 or smaller at 
$p^2 \sim 0.16$ GeV$^2$, below which $p^2$ the lattice evaluation was 
found unreliable \cite{Boucaud:2003xi}.}} at small momentum roughly 
as $p^4$ \cite{Boucaud:2003xi,Boucaud:2002fx}.

Some other, quite independent methods, also give the QCD 
running coupling vanishing at small $p^2$ \cite{VanAcoleyen:2002nd}
although not so fast as our form (\ref{ourAlpha_eff}).
Now, we want to make clear that we do not argue that our results 
are another indication that the running coupling indeed vanishes
as $p^2 \to 0$, because we are aware (as we comment in more detail 
below) that the behavior of our running coupling at very small $p^2$ 
is just an artifact of the way Eq. (\ref{ourAlpha_eff}) was derived.
Fortunately however, it turns out that the small-$p^2$ behavior
does not influence much our final, observable results (in 
contradistinction to intermediate $p^2$'s). To see all
this, let us discuss in detail the behavior of our form 
(\ref{ourAlpha_eff}) and especially the possible objections to it.
The first, less serious one is that 
$\alpha_{\mbox{\rm\scriptsize pe}}(p^2)$ ultimately hits the 
Landau pole as $p^2$ gets lower. However, we handle this
as in other phenomenological DS studies 
\cite{Jain:qh,Maris:1997tm,Maris:1999nt,Kekez:1996az,Klabucar:1997zi,Kekez:1998xr,Kekez:1998rw,Kekez:2001ph}, 
where 
this pole is shifted to timelike momenta in all logarithms:
$\ln(p^2/\Lqcd^2) \rightarrow \ln(x_0 + p^2/\Lqcd^2)$.
(Dynamical gluon mass can provide the physical reason 
for this \cite{Cornwall:1981zr};
i.e., $x_0 \propto m_A^2/\Lqcd^2 \sim 10$.)
For $\alpha_{\mbox{\rm\scriptsize pe}}(p^2)$ we use 
the two-loop expression used before by 
JM \cite{Jain:qh} and our earlier DS studies, e.g., Refs. 
\cite{Kekez:1996az,Klabucar:1997zi,Kekez:1998xr,Kekez:1998rw,Kekez:2001ph}. 
This means the infrared (IR) regulator (to which all results 
are almost totally insensitive) is $x_0 = 10$,
and $\Lqcd = 0.228$ GeV.
The parameters of $\alpha_{\mbox{\rm\scriptsize pe}}(p^2)$ 
are thereby fixed and do not belong among variable parameters
such as $C_A$ and $C_G$.

Back to the possible objections: 
the second,
in the present context the 
more serious one is that we cannot in advance give an 
argument that the factor $Z^{\Npert}(p^2) \, G^{\Npert}(p^2)^2$ 
in the proposed $\alpha_{\eff}(p^2)$ (\ref{ourAlpha_eff}) indeed 
approximates well nonperturbative contributions at low $p^2$
(say, $p^2 < 1$ GeV$^2$), but can only hope that our results 
to be calculated will provide an {\it a posteriori} 
justification for using it as low as $p^2 \sim 0.3$ GeV$^2$
[since Eq. (\ref{ourAlpha_eff}) takes appreciable values 
down to about $p^2 \sim 0.3$ GeV$^2$].
Of course, $Z^{\Npert}(p^2)$ and $G^{\Npert}(p^2)$ must 
be wrong in the limit $p^2 \rightarrow 0$, as the 
OPE-based results (\ref{KondoZ})-(\ref{gluonMass}) of Refs.
\cite{Lavelle:eg,Lavelle:xg,Ahlbach:ws,Lavelle:yh,Kondo:2001nq}
certainly fail in that limit. 
Thus, the extreme suppression for small $p^2$ is an unrealistic 
artifact of the proposed form (\ref{ourAlpha_eff}) when applied 
down to the $p^2 \rightarrow 0$ limit. Nevertheless, 
because of the integration measure in the integral equations
in DS calculations, integrands at these small $p^2$ 
do not contribute much, at least not to the quantities (such as 
$\langle {\bar q}q \rangle$ condensate, meson masses, decay 
constants and amplitudes) calculated in phenomenological DS analyses.
Hence, the form of $\alpha_{\eff}(p^2)$ at $p^2$ close to zero 
is not very important for the outcome of these phenomenological 
DS calculations.{\footnote{This is supported by, e.g., Fischer 
\cite{Fischer:2003zc}, who found that nearly all dynamically 
generated mass is produced by the integration strength above 
$p^2 = 0.25$ GeV$^2$.}}
This is 
because the most important for the success of phenomenological DS 
calculations seems the enhancement at somewhat higher values of 
$p^2$ - e.g., see the humps at $p^2 \sim 0.4$ to $0.6$
GeV$^2$ in the JM \cite{Jain:qh}
or MRT interaction \cite{Maris:1997tm,Maris:1999nt}
in Fig. \ref{Fig1}. 
Our $\alpha_{\eff}(p^2)$ (\ref{ourAlpha_eff}), the solid curve in 
Fig. \ref{Fig1}, 
exhibits such an enhancement centered around $p^2 \approx m_A^2/2$. 
This enhancement is readily understood when one notices that 
Eq. (\ref{ourAlpha_eff}) has four poles, 
\begin{eqnarray}
(p^2)_{1,2} &=& \frac{1}{2} \, \,
( \, \, m_A^2 \, \pm \, {\rm i} \, \sqrt{4 C_G - m_A^4} \, \, \, )~,
\\
(p^2)_{3,4} &=& \frac{1}{2} \, \,
( \, \, - m_A^2 \, \pm \, {\rm i} \, \sqrt{4 C_A - m_A^4} \, \, \, )~,
\label{poles}
\end{eqnarray}
in the complex $p^2$ plane.
For $\min\{ C_G, C_A\} > m_A^4/4$ there are no poles
on the real axis, but 
saddles between two complex conjugated poles. For 
the DS studies, which are almost exclusively carried out in Euclidean
space, spacelike $p^2$, i.e., $p^2 > 0$ is the relevant domain and is
thus pictured in Fig. \ref{Fig1}. There, the maximum of 
$\alpha_{\eff}(p^2)$ (\ref{ourAlpha_eff}) at the real axis is at 
$p^2 \approx m_A^2/2$, i.e., the real part of
its {\it double} poles $(p^2)_{1,2}$ coming from $G^{\Npert}(p^2)^2$. 
The height and the width of the peak is influenced by both $C_G$ 
and $m_A$. The enhancement of 
$\alpha_{\eff}(p^2)$ (\ref{ourAlpha_eff}) is thus
determined by $\langle A^2 \rangle$ through
Eq. (\ref{gluonMass}), and by the manner this condensate contributes 
to the ghost renormalization function.

We are thus motivated to use this form (\ref{ourAlpha_eff}) of 
$\alpha_{\eff}(p^2)$ for all $p^2$ to test its success in 
DS calculations. We are aware of the shortcomings 
due to its oversimplified character, but its study helps to answer 
whether the $\langle A^2 \rangle$ condensate, which has recently 
attracted so much attention, may be important for the
enhancement of the effective interaction needed for successful 
DS phenomenology.  
The results presented below indicate that the
$\langle A^2 \rangle$ condensate may indeed provide an
important mechanism not considered so far.

\section{End results with discussion and conclusion}

We solved the gap and BS equations (\ref{schematically}) 
for quark pro-\\pagators 
\begin{equation}
S(p) \, = \,  
\frac{1}{{\rm i}\gamma \cdot p \, {\cal A}(p^2) + {\cal B}(p^2) }
  \, \equiv \, 
\frac{ {\cal A}(p^2)^{-1}}{ {\rm i} \gamma \cdot p + M(p^2) }
\label{FULLqPropag}
\end{equation}
and for pseudoscalar meson $q\bar q$ ($q=u,d,s$) 
bound-state vertices ($\Gamma_{\tt M}$) in the same way as in 
our previous phenomenological DS studies, e.g., 
Refs. \cite{Kekez:1996az,Klabucar:1997zi,Kekez:1998xr,Kekez:1998rw,Kekez:2001ph}. 
This essentially means as in the JM approach \cite{Jain:qh},
except that instead of JM's $\alpha_{\eff}(p^2)$, 
Eq. (\ref{ourAlpha_eff}) is employed in the RLA interaction (\ref{RLAkernel}). 
We can thus immediately present the results because we can refer to 
Refs. \cite{Jain:qh,Kekez:1996az,Klabucar:1997zi,Kekez:1998xr,Kekez:1998rw,Kekez:2001ph} 
for all calculational details, such as procedures 
for solving DS and BS equations, all model details, 
as well as explicit expressions for calculated quantities
and inputs such as the aforementioned IR-regularized 
$\alpha_{\pert}(p^2)$.

\begin{figure}[t]
\centerline{\includegraphics[height=66mm,angle=0]{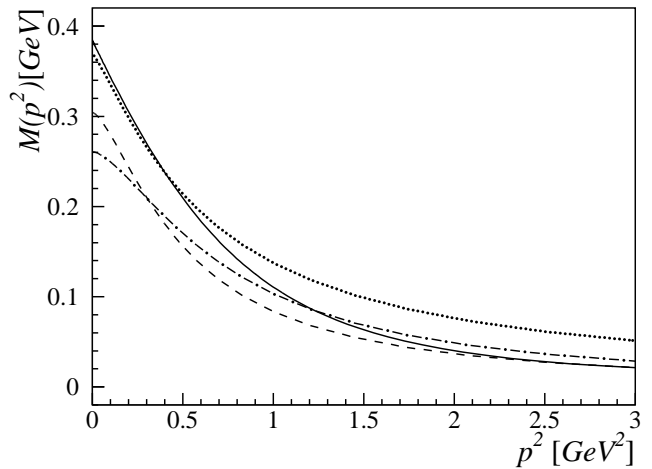}}
\caption{The momentum dependence of the dynamically generated 
quark mass $M(p^2)$ for $u$ and $d$ quarks. 
The solid curve is our result for the parameters giving the
second line of Table I, but our $M(p^2)$ depends in fact very little
on the small explicit chiral symmetry breaking mass parameters
${\widetilde m}_u$ and ${\widetilde m}_d$ of the very light 
$u$ and $d$ quarks as long as their values are at all realistic. 
The dotted curve is the 
{\it ab initio} DS result \cite{Fischer:2003rp,Alkofer:2002aa}.
The short-dashed curve is the $M(p^2)$-fit of Ref. \cite{Zhang:2003fa} 
to the extrapolation of their lattice data to the chiral limit.
The dash-dotted curve is the similar result from another 
lattice calculation, namely the fit of $M(p^2)$ 
from Fig. 14 of Ref. \cite{Bowman:2002bm}. 
}
\label{FigM}
\end{figure}

These calculations show 
that the initial $m_A, C_A$ and $C_G$ estimates
motivated and given above, need only slight (a few \%)
modifications to provide a very good description of the light
pseudoscalar sector. Concretely, now we will [both in the 
chiral limit ($\chi\lim$) and realistically away from it]
quote only results obtained for the parameter set
\begin{equation}
C_A = (0.6060 \,\, {\rm GeV})^4 = C_G \,\, , \,\,
m_A = 0.8402 \,\, {\rm GeV} \, ,
\label{StandardParameterSet-new}
\end{equation}
while a broader investigation of parameter dependence 
shows the following. 
{\it i)} The results are only weakly sensitive to
moderate variations (up to the factors of 2 to 1/2) of $C_A$.
{\it ii)} Contrary to that, the results are very sensitive 
to $m_A$ and $C_G$, since they determine the peak of our
$\alpha_{\eff}(p^2)$ (\ref{ourAlpha_eff}). However,
between $C_G^{\minimal} \sim (0.6$ GeV$)^4$
and $C_G^{\maximal} \sim (0.9$ GeV$)^4$
there are many pairs of these quantities
which give fits comparable 
(within a percent) to that
resulting from the values (\ref{StandardParameterSet-new}),
as long as they approximately satisfy the linear relation
\begin{equation}
( C_G )^{1/4} = 0.7742 \, m_A - 0.0442 \,\, {\rm GeV}~.
\label{CG-mArelation}
\end{equation}
Thus, the two parameters ruling the strength
of $\alpha_{\eff}(p^2)$ are not independent.

\begin{table}[b]
\vspace*{1ex}
\begin{tabular}{|c|c|c|l|c|}
\hline
 $M_{\pi^0}$  & $ ~f_{\pi^+}$  & 
$M_{K^{+}}$  & $\, ~~f_{K^+}$  &
$T_{\pi^0}^{\gamma\gamma}$ [GeV$^{-1}$] \\
\hline
    \, 136.17 \, &  93.0  & \, 516.28 \, & \, 112.5  & $ 0.256 $  \\
\hline
     134.96 &  92.9  & 494.92 & \, 111.5  & $ 0.256 $ \\
\hline
 $134.98 $ & $  92.4\pm 0.3  $ & 493.68 & $ 113.0\pm 1.0  $  &
                                $ 0.274 \pm 0.010 $  \\
\hline
\end{tabular}
\vspace*{1ex}
\caption{The masses and decay constants of pions and kaons and
$\pi^0\to\gamma\gamma$ decay amplitude $T_{\pi^0}^{\gamma\gamma}$,
for the parameter values (\ref{StandardParameterSet-new})
and our $\alpha_{\eff}(p^2)$ (\ref{ourAlpha_eff}).
The JM \cite{Jain:qh} quark bare masses
${\widetilde m}_u = {\widetilde m}_d = 3.1 ~\MeV
 \, , \, {\widetilde m}_s = 73 ~\MeV$ give the first line.
The second line are the results obtained with
${\widetilde m}_u = {\widetilde m}_d = 3.046 ~\MeV
 \, , \,
{\widetilde m}_s = 67.70 ~\MeV$.
The last line gives the experimental values. (The distinction
between neutral and charged mesons applies only to this line,
as we calculate in the isosymmetric limit.)
Everything is in MeV except $T_{\pi^0}^{\gamma\gamma}$.  }
\label{tab:massiveResults}
\end{table}

Already the chiral-limit results are very satisfactory:
$f_{\pi^\pm} = f_{\pi^0} \equiv f_\pi = 90.5$ MeV,
the $\pi^0 \to \gamma\gamma$ chiral-limit amplitude
$T_{\pi^0}^{\gamma\gamma}(\chi\lim)\equiv 1/(4\pi^2 f_\pi)
 = 0.280$ GeV$^{-1}$ and the chiral condensate
$\langle\bar{q}q\rangle = (-217 \, {\rm MeV})^3$.
We also get the correct QCD chiral-limit behavior: massless $q\bar q$ 
pseudoscalars and satisfied (within $\sim 4$\%) GMOR.

In the chiral limit, where  
the quark mass is purely dynamically generated since
the bare (and current) quark 
masses vanish, the only parameters are $m_A, C_A$ and $C_G$.
Away from the chiral limit, chiral symmetry is explicitly broken 
by the nonvanishing bare mass parameters ${\widetilde m}_q$ 
of light quarks ($q=u,d,s$) entering the quark-propagator 
gap equation and the $q\bar q$ BS equation. 
For the very light quarks $u$ and $d$, 
the dynamically generated quark masses of $u$- and $d$-quarks 
away from the chiral limit are practically the same as in the 
chiral limit.
In Fig. \ref{FigM}, 
the solid curve depicts our results for the momentum dependence 
of these dynamical masses 
$M(p^2) \equiv {\cal B}(p^2)/{\cal A}(p^2)$
of $u$- and $d$-quarks (in the isosymmetric limit). 
Fig. \ref{FigM} also presents the lattice results for 
the $u$- and $d$-quark dynamical masses of 
Refs. \cite{Bowman:2002bm,Zhang:2003fa} 
which give their fits to the lattice data points
in terms of analytic, closed-form expressions. 
Our results for the dynamical quark mass $M(p^2)$ agree very well with 
both the {\it ab initio} DS results \cite{Fischer:2003rp,Alkofer:2002aa}
and the lattice results \cite{Bowman:2002bm,Zhang:2003fa}.

As the first attempt to depart realistically from 
the chiral limit, we adopt without change  
the ${\widetilde m}_q$ values obtained earlier
by JM \cite{Jain:qh} in a very broad DS fit of
the meson phenomenology (with {\it their}
$\alpha_{\eff}$ \cite{Jain:qh}), i.e.,
${\widetilde m}_u = {\widetilde m}_d = 3.1$ MeV and 
${\widetilde m}_s = 73$ MeV.
Already the corresponding results for the masses and decay 
constants of pions and kaons and the $\pi^0 \to \gamma\gamma$ 
decay amplitude $T_{\pi^0}^{\gamma\gamma}$, given in the 
first line of Table \ref{tab:massiveResults}, show a very good
agreement with experimental values, except for the kaon mass.
The second line shows that just a slight re-adjustment of the 
quark masses, to ${\widetilde m}_u = {\widetilde m}_d = 3.046 
~\MeV \, , \, {\widetilde m}_s = 67.70 ~\MeV$, is enough to 
get an almost perfect fit to the pion and kaon masses.

The presented results allow
us to conclude that the dimension 2 gluon condensate 
$\langle A^2 \rangle$ provides an enhanced effective
interaction $\alpha_{\eff}(p^2)$ which leads to a 
sufficiently strong D$\chi$SB, pions and kaons as 
(quasi-)Goldstone bosons of QCD, and successful DS 
phenomenology at least in the light sector of 
pseudoscalar mesons. This opens the possibility 
that instead of modeling $\alpha_{\eff}(p^2)$,
its enhancement at intermediate $p^2$ may be 
understood in terms of gluon condensates.

\section*{Acknowledgment}
\noindent The authors gratefully acknowledge the support of
the Croatian Ministry of Science and Technology contracts
0119261 and 0098011.


\begin{thebibliography}{100}

%\cite{Alkofer:2000wg}
\bibitem{Alkofer:2000wg}
R.~Alkofer and L.~von Smekal,
%``The infrared behavior of QCD Green's functions: Confinement, dynamical
%symmetry breaking, and hadrons as relativistic bound states,''
Phys.\ Rept.\  {\bf 353}, 281 (2001)
[arXiv:hep-ph/0007355].
%%CITATION = HEP-PH 0007355;%%


%\cite{Fischer:2003rp}
\bibitem{Fischer:2003rp}
C.~S.~Fischer and R.~Alkofer,
%``Non-perturbative propagators, running coupling and dynamical quark mass of
%Landau gauge QCD,''
Phys.\ Rev.\ D {\bf 67}, 094020 (2003)
[arXiv:hep-ph/0301094].
%%CITATION = HEP-PH 0301094;%%


%\cite{Maris:2003vk}
\bibitem{Maris:2003vk}
P.~Maris and C.D.~Roberts,
%``Dyson-Schwinger equations: A tool for hadron physics,''
Int.\ J.\ Mod.\ Phys.\ E {\bf 12}, 297 (2003)
[arXiv:nucl-th/0301049].
%%CITATION = NUCL-TH 0301049;%%


%\cite{Jain:qh}
\bibitem{Jain:qh}
P.~Jain and H.~J.~Munczek,
%``Q Anti-Q Bound States In The Bethe-Salpeter Formalism,''
Phys.\ Rev.\ D {\bf 48}, 5403 (1993)
[arXiv:hep-ph/9307221].
%%CITATION = HEP-PH 9307221;%%


%\cite{Maris:1997tm}
\bibitem{Maris:1997tm}
P.~Maris and C.~D.~Roberts,
%``pi and K meson Bethe-Salpeter amplitudes,''
Phys.\ Rev.\ C {\bf 56}, 3369 (1997)
[arXiv:nucl-th/9708029].
%%CITATION = NUCL-TH 9708029;%%


%\cite{Maris:1999nt}
\bibitem{Maris:1999nt}
P.~Maris and P.~C.~Tandy,
%``Bethe-Salpeter study of vector meson masses and decay constants,''
Phys.\ Rev.\ C {\bf 60}, 055214 (1999)
[arXiv:nucl-th/9905056].
%%CITATION = NUCL-TH 9905056;%%


%\cite{Holl:2003dq}
\bibitem{Holl:2003dq}
A.~Holl, A.~Krassnigg and C.~D.~Roberts,
%``DSEs and pseudoscalar mesons: an aperc,u,''
arXiv:nucl-th/0311033.
%%CITATION = NUCL-TH 0311033;%%


%\cite{Bender:1996bb}
\bibitem{Bender:1996bb}
A.~Bender, C.~D.~Roberts and L.~Von Smekal,
%``Goldstone Theorem and Diquark Confinement Beyond Rainbow-Ladder
%Approximation,''
Phys.\ Lett.\ B {\bf 380}, 7 (1996)
[arXiv:nucl-th/9602012].
%%CITATION = NUCL-TH 9602012;%%


%\cite{Bender:2002as}
\bibitem{Bender:2002as}
A.~Bender, W.~Detmold, C.~D.~Roberts and A.~W.~Thomas,
%``Bethe-Salpeter equation and a nonperturbative quark gluon vertex,''
Phys.\ Rev.\ C {\bf 65}, 065203 (2002)
[arXiv:nucl-th/0202082].
%%CITATION = NUCL-TH 0202082;%%


%\cite{Maris:1997hd}
\bibitem{Maris:1997hd}
P.~Maris, C.~D.~Roberts and P.~C.~Tandy,
%``Pion mass and decay constant,''
Phys.\ Lett.\ B {\bf 420}, 267 (1998)
[arXiv:nucl-th/9707003].
%%CITATION = NUCL-TH 9707003;%%


%\cite{Maris:2002mt}
\bibitem{Maris:2002mt}
P.~Maris, A.~Raya, C.~D.~Roberts and S.~M.~Schmidt,
%``Facets of confinement and dynamical chiral symmetry breaking,''
Eur.\ Phys.\ J.\ A {\bf 18}, 231 (2003)
[arXiv:nucl-th/0208071].
%%CITATION = NUCL-TH 0208071;%%


%\cite{Bloch:2002we}
\bibitem{Bloch:2002we}
J.~C.~R.~Bloch, A.~Cucchieri, K.~Langfeld and T.~Mendes,
%``Running coupling constant and propagators in SU(2) Landau gauge,''
Nucl.\ Phys.\ Proc.\ Suppl.\  {\bf 119}, 736 (2003)
[arXiv:hep-lat/0209040].
%%CITATION = HEP-LAT 0209040;%%


%\cite{Boucaud:2000nd}
\bibitem{Boucaud:2000nd}
P.~Boucaud, A.~Le Yaouanc, J.~P.~Leroy, J.~Micheli, O.~Pene and J.~Rodriguez-Quintero,
%``Consistent OPE description of gluon two point and three point Green
%function?,''
Phys.\ Lett.\ B {\bf 493}, 315 (2000)
[arXiv:hep-ph/0008043].
%%CITATION = HEP-PH 0008043;%%


%\cite{Gubarev:2000eu}
\bibitem{Gubarev:2000eu}
F.~V.~Gubarev, L.~Stodolsky and V.~I.~Zakharov,
%``On the significance of the quantity A**2,''
Phys.\ Rev.\ Lett.\  {\bf 86}, 2220 (2001)
[arXiv:hep-ph/0010057].
%%CITATION = HEP-PH 0010057;%%


%\cite{Gubarev:2000nz}
\bibitem{Gubarev:2000nz}
F.~V.~Gubarev and V.~I.~Zakharov,
%``On the emerging phenomenology of <(A(a)(mu))**2(min)>,''
Phys.\ Lett.\ B {\bf 501}, 28 (2001)
[arXiv:hep-ph/0010096].
%%CITATION = HEP-PH 0010096;%%


%\cite{Kondo:2001nq}
\bibitem{Kondo:2001nq}
K.~I.~Kondo,
%``Vacuum condensate of mass dimension 2 as the origin of mass gap and  quark
%confinement,''
Phys.\ Lett.\ B {\bf 514}, 335 (2001)
[arXiv:hep-th/0105299].
%%CITATION = HEP-TH 0105299;%%


%\cite{Celenza:1986th}
\bibitem{Celenza:1986th}
A very early work considering the $\langle A^2 \rangle$
gluon condensate seriously, is
L.~S.~Celenza and C.~M.~Shakin,
%``Description Of The Gluon Condensate,''
Phys.\ Rev.\ D {\bf 34}, 1591 (1986).
%%CITATION = PHRVA,D34,1591;%%


%\cite{Lavelle:eg}
\bibitem{Lavelle:eg}
M.~J.~Lavelle and M.~Schaden,
%``Propagators And Condensates In QCD,''
Phys.\ Lett.\ B {\bf 208}, 297 (1988).
%%CITATION = PHLTA,B208,297;%%


%\cite{Lavelle:xg}
\bibitem{Lavelle:xg}
M.~Lavelle and M.~Schaden,
%``The Gluon Condensate And The Effective Potential In QCD,''
Phys.\ Lett.\ B {\bf 246}, 487 (1990).
%%CITATION = PHLTA,B246,487;%%



%\cite{Ahlbach:ws}
\bibitem{Ahlbach:ws}
J.~Ahlbach, M.~Lavelle, M.~Schaden and A.~Streibl,
%``Propagators And Four-Dimensional Condensates In Pure QCD,''
Phys.\ Lett.\ B {\bf 275}, 124 (1992).
%%CITATION = PHLTA,B275,124;%%


%\cite{Lavelle:yh}
\bibitem{Lavelle:yh}
M.~Lavelle and M.~Oleszczuk,
%``The Operator Product Expansion Of The QCD Propagators,''
Mod.\ Phys.\ Lett.\ A {\bf 7}, 3617 (1992).
%%CITATION = MPLAE,A7,3617;%%



%\cite{Kondo:2001tm}
\bibitem{Kondo:2001tm}
K.~I.~Kondo, T.~Murakami, T.~Shinohara and T.~Imai,
%``Renormalizing a BRST-invariant composite operator of mass dimension 2  in
%Yang-Mills theory,''
Phys.\ Rev.\ D {\bf 65}, 085034 (2002)
[arXiv:hep-th/0111256].
%%CITATION = HEP-TH 0111256;%%


%\cite{Kondo:2002xn}
\bibitem{Kondo:2002xn}
K.~I.~Kondo and T.~Imai,
%``A confining string theory derivable from Yang-Mills theory due to a  novel
%vacuum condensate,''
arXiv:hep-th/0206173.
%%CITATION = HEP-TH 0206173;%%



%\cite{Dudal:2002xe}
\bibitem{Dudal:2002xe}
D.~Dudal and H.~Verschelde,
%``On ghost condensation, mass generation and Abelian dominance in the maximal Abelian gauge,''
J.\ Phys.\ A {\bf 36}, 8507 (2003)
[arXiv:hep-th/0209025].
%%CITATION = HEP-TH 0209025;%%



%\cite{Shifman:bx}
\bibitem{Shifman:bx}
M.~A.~Shifman, A.~I.~Vainshtein and V.~I.~Zakharov,
%``QCD And Resonance Physics. Sum Rules,''
Nucl.\ Phys.\ B {\bf 147}, 385 (1979); {\it ibid.} 448 (1979).
%%CITATION = NUPHA,B147,385;%%


%\cite{Lavelle:ve}
\bibitem{Lavelle:ve}
M.~Lavelle,
%``Gauge Invariant Effective Gluon Mass From The Operator Product Expansion,''
Phys.\ Rev.\ D {\bf 44}, 26 (1991).
%%CITATION = PHRVA,D44,26;%%


%\cite{Ioffe:2002be}
\bibitem{Ioffe:2002be}
B.~L.~Ioffe and K.~N.~Zyablyuk,
%``Gluon condensate in charmonium sum rules with 3-loop corrections,''
Eur.\ Phys.\ J.\ C {\bf 27}, 229 (2003)
[arXiv:hep-ph/0207183].
%%CITATION = HEP-PH 0207183;%%


%\cite{Boucaud:2001st}
\bibitem{Boucaud:2001st}
P.~Boucaud, A.~Le Yaouanc, J.~P.~Leroy, J.~Micheli, O.~Pene and J.~Rodriguez-Quintero,
%``Testing Landau gauge OPE on the lattice with a condensate,''
Phys.\ Rev.\ D {\bf 63}, 114003 (2001)
[arXiv:hep-ph/0101302].
%%CITATION = HEP-PH 0101302;%%



%\cite{Fischer:2002hn}
\bibitem{Fischer:2002hn}
C.~S.~Fischer and R.~Alkofer,
%``Infrared exponents and running coupling of SU(N) Yang-Mills theories,''
Phys.\ Lett.\ B {\bf 536}, 177 (2002)
[arXiv:hep-ph/0202202].
%%CITATION = HEP-PH 0202202;%%



%\cite{Alkofer:2002aa}
\bibitem{Alkofer:2002aa}
R.~Alkofer, C.~S.~Fischer and L.~von Smekal,
%``Infrared exponents and the running coupling of Landau gauge QCD and their
%relation to confinement,''
Eur.\ Phys.\ J.\ A {\bf 17}, 773 (2003)
[arXiv:hep-ph/0209366].
%%CITATION = HEP-PH 0209366;%%



%\cite{Leinweber:1998uu}
\bibitem{Leinweber:1998uu}
D.~B.~Leinweber, J.~I.~Skullerud, A.~G.~Williams and C.~Parrinello  [UKQCD
                  Collaboration],
%``Asymptotic scaling and infrared behavior of the gluon propagator,''
Phys.\ Rev.\ D {\bf 60}, 094507 (1999)
[Erratum-ibid.\ D {\bf 61}, 079901 (2000)]
[arXiv:hep-lat/9811027].
%%CITATION = HEP-LAT 9811027;%%


%\cite{Iida:2003xe}
\bibitem{Iida:2003xe}
H.~Iida, M.~Oka and H.~Suganuma,
%``Dynamical chiral-symmetry breaking at t = 0 and t not equal 0 in the
%Schwinger-Dyson equation with lattice QCD data,''
arXiv:hep-ph/0312328.
%%CITATION = HEP-PH 0312328;%%


\bibitem{BBLWZ00} F.~D.~R.~Bonnet, P.~O.~Bowman, D.~B.~Leinweber, A.~G.~Williams 
and J.~M.~Zanotti, Phys. Rev. D{\bf 64}, 034501 (2001) [arXiv:hep-lat/0101013];
F.~D.~R.~Bonnet, P.~O.~Bowman, D.~B.~Leinweber and A.~G.~Williams, Phys. Rev. 
D {\bf 62}, 051501 (2000) [arXiv:hep-lat/0002020].



\bibitem{SaoCarlos}
T\"ubingen lattice data are also supported 
\cite{Bloch:2002we,Bloch:2003sk} by S\~ao Carlos lattice data.


%\cite{Bloch:2003sk}
\bibitem{Bloch:2003sk}
J.~C.~R.~Bloch, A.~Cucchieri, K.~Langfeld and T.~Mendes,
%``Propagators and running coupling from SU(2) lattice gauge theory,''
Nucl.\ Phys.\ B {\bf 687}, 76 (2004)
[arXiv:hep-lat/0312036].
%%CITATION = HEP-LAT 0312036;%%


%\cite{Suman:1995zg}
\bibitem{Suman:1995zg}
H.~Suman and K.~Schilling,
%``First Lattice Study of Ghost Propagators in SU(2) and SU(3) Gauge Theories,''
Phys.\ Lett.\ B {\bf 373}, 314 (1996)
[arXiv:hep-lat/9512003].
%%CITATION = HEP-LAT 9512003;%%


%\cite{Suman:be}
\bibitem{Suman:be}
H.~Suman and K.~Schilling,
%``Dipole Structure Of Ghost Propagators In Gauge Theories,''
Nucl.\ Phys.\ Proc.\ Suppl.\  {\bf 53}, 850 (1997).
%%CITATION = NUPHZ,53,850;%%


%\cite{Boucaud:2003xi}
\bibitem{Boucaud:2003xi}
P.~Boucaud, F.~De Soto, A.~Le Yaouanc, J.~P.~Leroy, J.~Micheli, 
O.~Pene and J.~Rodriguez-Quintero,
%``Evidences for instantons effects in Landau lattice Green functions,''
arXiv:hep-ph/0312332.
%%CITATION = HEP-PH 0312332;%%


%\cite{Boucaud:2002fx}
\bibitem{Boucaud:2002fx}
P.~Boucaud, F.~De Soto, A.~Le Yaouanc, J.~P.~Leroy, J.~Micheli, 
H. Moutarde, O.~Pene and J.~Rodriguez-Quintero, 
%``The strong coupling constant at small momentum as an instanton detector,''
JHEP {\bf 0304}, 005 (2003)
[arXiv:hep-ph/0212192].
%%CITATION = HEP-PH 0212192;%%


%\cite{VanAcoleyen:2002nd}
\bibitem{VanAcoleyen:2002nd}
K.~Van Acoleyen and H.~Verschelde,
%``Avoiding the Landau-pole in perturbative QCD,''
Phys.\ Rev.\ D {\bf 66}, 125012 (2002)
[arXiv:hep-ph/0203211].
%%CITATION = HEP-PH 0203211;%%


%\cite{Kekez:1996az}
\bibitem{Kekez:1996az}
D.~Kekez and D.~Klabu\v{c}ar,
%``Two--photon processes of pseudoscalar mesons in a Bethe--Salpeter approach,''
Phys.\ Lett.\ B {\bf 387}, 14 (1996)
[arXiv:hep-ph/9605219].
%%CITATION = HEP-PH 9605219;%%


%\cite{Klabucar:1997zi}
\bibitem{Klabucar:1997zi}
D.~Klabu\v{c}ar and D.~Kekez,
%``eta and eta' at the limits of applicability of a coupled  Schwinger-Dyson and
%Bethe-Salpeter approach in the ladder  approximation,''
Phys.\ Rev.\ D {\bf 58}, 096003 (1998)
[arXiv:hep-ph/9710206].
%%CITATION = HEP-PH 9710206;%%


%\cite{Kekez:1998xr}
\bibitem{Kekez:1998xr}
D.~Kekez, B.~Bistrovi\'{c} and D.~Klabu\v{c}ar,
%``Application of Jain and Munczek's bound-state approach to gamma gamma
%processes of pi0, eta/c and eta/b,''
Int.\ J.\ Mod.\ Phys.\ A {\bf 14}, 161 (1999)
[arXiv:hep-ph/9809245].
%%CITATION = HEP-PH 9809245;%%


%\cite{Kekez:1998rw}
\bibitem{Kekez:1998rw}
D.~Kekez and D.~Klabu\v{c}ar,
%``gamma* gamma $\to$ pi0 transition and asymptotics of gamma* gamma and  
%gamma* gamma* transitions of other unflavored pseudoscalar mesons,''
Phys.\ Lett.\ B {\bf 457}, 359 (1999)
[arXiv:hep-ph/9812495].
%%CITATION = HEP-PH 9812495;%%


%\cite{Kekez:2001ph}
\bibitem{Kekez:2001ph}
D.~Kekez and D.~Klabu\v{c}ar,
%``eta and eta' in a coupled Schwinger-Dyson and Bethe-Salpeter approach:  The
%gamma* gamma transition form factors,''
Phys.\ Rev.\ D {\bf 65}, 057901 (2002)
[arXiv:hep-ph/0110019].
%%CITATION = HEP-PH 0110019;%%



%\cite{Cornwall:1981zr}
\bibitem{Cornwall:1981zr}
J.~M.~Cornwall,
%``Dynamical Mass Generation In Continuum QCD,''
Phys.\ Rev.\ D {\bf 26}, 1453 (1982).
%%CITATION = PHRVA,D26,1453;%%


%\cite{Fischer:2003zc}
\bibitem{Fischer:2003zc}
C.~S.~Fischer,
%``Non-perturbative propagators, running coupling and dynamical mass  generation
%in ghost - antighost symmetric gauges in QCD,''
arXiv:hep-ph/0304233.
%%CITATION = HEP-PH 0304233;%%


%\cite{Zhang:2003fa}
\bibitem{Zhang:2003fa}
J.~B.~Zhang, P.~O.~Bowman, D.~B.~Leinweber, A.~G.~Williams and F.~D.~R.~Bonnet
                  [CSSM Lattice collaboration],
%``Scaling behavior of the overlap quark propagator in Landau gauge,''
Phys.\ Rev.\ D {\bf 70}, 034505 (2004)
[arXiv:hep-lat/0301018].
%%CITATION = HEP-LAT 0301018;%%


%\cite{Bowman:2002bm}
\bibitem{Bowman:2002bm}
P.~O.~Bowman, U.~M.~Heller and A.~G.~Williams,
%``Lattice quark propagator with staggered quarks in Landau and Laplacian
%gauges,''
Phys.\ Rev.\ D {\bf 66}, 014505 (2002)
[arXiv:hep-lat/0203001].
%%CITATION = HEP-LAT 0203001;%%


\end{thebibliography}
\end{document}